\documentclass[prd,aps,showpacs,nofootinbib,preprint,eqsecnum]{revtex4}
\usepackage{amsmath}
\usepackage{amssymb}
\usepackage{amsfonts}
\usepackage{graphicx,bm}
\usepackage{dcolumn}
\usepackage{color,amsxtra}
\usepackage{epsf}
\usepackage{enumerate}
\usepackage{hhline}
\usepackage{array}
\usepackage{tabularx}
\usepackage{subfigure}
\usepackage{fancyhdr}
\usepackage{mathrsfs}

\usepackage{amsmath}
\usepackage{amsthm}
\usepackage{amssymb}
\usepackage{amscd}
\usepackage[british]{babel}
\usepackage{babel}
\usepackage{graphicx}
\usepackage{psfrag}
\usepackage{epsfig}
\usepackage{rotating}
\usepackage{times}

\newcommand{\Eqn}[1]{&\hspace{-0.2em}#1\hspace{-0.2em}&}

\def\Vec#1{\mbox{\boldmath $#1$}}

\theoremstyle{plain}

\newtheorem{remark}{Remark}[section]

\newcommand{\boxend}{\flushright{$\Box$}}

\newcommand{\Z}{{\mathbb Z}}               
\newcommand{\R}{{\mathbb R}}               

\newcommand{\e}{\epsilon}

\begin{document}


\title{Future singularities and Teleparallelism
in Loop Quantum Cosmology}
\author{
Kazuharu Bamba$^{1, }$\footnote{
E-mail address: bamba@kmi.nagoya-u.ac.jp},
Jaume de Haro$^{2, }$\footnote{E-mail address: jaime.haro@upc.edu}
and
Sergei D. Odintsov$^{3, 4, 5, 6, }$\footnote{
E-mail address: odintsov@ieec.uab.es}
}
\affiliation{
$^1$Kobayashi-Maskawa Institute for the Origin of Particles and the
Universe,
Nagoya University, Nagoya 464-8602, Japan\\
$^2$Departament de Matem\`atica Aplicada I, Universitat
Polit\`ecnica de Catalunya, Diagonal 647, 08028 Barcelona, Spain \\
$^3$Dept. Gen. and Theor. Phys. and Eurasian International Center for
Theoretical Physics, Eurasian National University, Astana 010008, Kazakhstan\\
$^4$Instituci\`{o} Catalana de Recerca i Estudis Avan\c{c}ats (ICREA),
Barcelona, Spain\\
$^5$Institut de Ciencies de l'Espai (CSIC-IEEC),
Campus UAB, Facultat de Ciencies, Torre C5-Par-2a pl, E-08193 Bellaterra
(Barcelona), Spain\\
$^6$Tomsk State Pedagogical University, Tomsk, Russia
}



\begin{abstract}

We demonstrate how holonomy corrections in loop quantum cosmology (LQC)
prevent the Big Rip singularity by
introducing a quadratic modification in terms of the energy density $\rho$
in the Friedmann equation in the Friedmann-Lema\^{i}tre-Robertson-Walker
(FLRW) space-time in a consistent and useful way.
In addition,
we investigate whether other kind of singularities like Type II,III and IV
singularities
survive or are avoided in LQC
when the universe is filled by a barotropic fluid with the
state equation $P=-\rho-f(\rho)$, where $P$ is the pressure and
$f(\rho)$ a function of $\rho$.
It is shown that
the Little Rip cosmology does not happen in LQC.
Nevertheless, the occurrence of the Pseudo-Rip cosmology, in which the phantom
universe approaches the de Sitter one asymptotically, is established, and
the corresponding example is presented.
It is interesting that the disintegration of
bound structures in the Pseudo-Rip cosmology in LQC always
takes more time than that in Einstein cosmology.
Our investigation on future singularities is generalized to that
in modified teleparallel gravity,
where LQC and Brane Cosmology in the Randall-Sundrum scenario
are the best examples. It is remarkable that $F(T)$ gravity
may lead to all the kinds of future singularities including Little
Rip.

\end{abstract}


\maketitle



\section{Introduction}

The accelerated expansion of the present universe
has been supported by
various cosmological observations, for example,
Type Ia Supernovae~\cite{SN1},
baryon acoustic oscillations (BAO)~\cite{Eisenstein:2005su},
large scale structure (LSS)~\cite{LSS},
cosmic microwave background (CMB) radiation~\cite{WMAP, Komatsu:2008hk, k11},
and effects of weak lensing~\cite{Jain:2003tba}.
As representative procedures to explain the late time acceleration,
the first is to assume the existence of
an unknown matter called ``dark energy'' in general relativity,
  which has negative pressure
(for recent reviews, e.g., see~\cite{Caldwell:2009ix, Maartens:2010ar,
Li:2011sd, Kunz:2012aw, Bamba:2012cp}).
The second is to modify gravity, the simplest way of which is
$F(R)$ gravity
(for recent reviews, for example, see~\cite{Review-Nojiri-Odintsov,
Book-Capozziello-Faraoni, Clifton:2011jh, Capozziello:2011et, Harko:2012ar}).


In general, dark energy can be assumed to be a perfect fluid with
the  equation of state $P=\rho-f(\rho)$, which realizes
the current cosmic acceleration.
Moreover, the Wilkinson Microwave Anisotropy Probe (WMAP) observations
indicate that the central value of the equation of state (EoS) is given by
$w \equiv P/\rho \cong-1.10$~\cite{k11}.
This means that our universe would be dominated by ``phantom energy''
($f(\rho)\geq 0$, i.e., $w < -1$).
However, the classical solutions of general relativity
for a Friedmann-Lema\^{i}tre-Robertson-Walker (FLRW)
model containing dark energy lead, in general,
to future singularities~\cite{ckw03, b04, not05, mss90, s00}
such as the Big Rip and future Sudden singularities.
In Ref.~\cite{not05}, the finite-time future singularities have been
classified into four types.

Recently, the possibility to avoid these future singularities
has been studied in the literature, by using different approaches to
quantum cosmology like loop quantum cosmology (LQC)~\cite{Bojowald:2001xe,
bo02a, Ashtekar:2005qt, a07, Ashtekar:2007em}
(for a recent example of a related study on LQC,
a super acceleration in LQC and its possible phase transitions, i.e.,
the crossing of the phantom divide line $w=-1$ has been discussed in
Ref.~\cite{Sadjadi:2012wg}),
semiclassical gravity, modified gravity,
brane cosmology, and etc
(for reviews on LQC, see, e.g.,~\cite{Thiemann:2002nj, abl03, Bojowald:2006da,
t01, Bojowald:2008zzb, as11, Bojowald:2012xy}).

Moreover, as an alternative gravitational theory to general relativity,
``teleparallelism''~\cite{Hehl:1976kj, hs79, Flanagan:2007dc, g10}
has recently been considered.
It is constructed by using the Weitzenb\"{o}ck connection.
Hence, the action is described by the torsion scalar $T$ and
not the scalar curvature $R$ as in general relativity
formulated with the Levi-Civita connection.
It is known that
the modified teleparallel gravity, so-called $F(T)$ gravity,
can realize both inflation~\cite{Inflation-F-F}
and the late-time cosmic
acceleration~\cite{bf09, Linder:2010py, BGL-Comment}\footnote{
If there exists a scalar field with a non-minimal coupling to
the torsion scalar $T$ even in pure teleparallelism and not in $F(T)$ gravity,
the late-time cosmic acceleration can be realized~\cite{TDE}. Moreover,
the generation of large-scale magnetic fields from inflation
in the framework of pure teleparallelism has been investigated
in Ref.~\cite{Bamba:2012mi}.}
(for more detailed references on $F(T)$ gravity,
see, e.g.,~\cite{Bamba:2012cp}).
In Ref.~\cite{bmno12}, models of $F(T)$ gravity in which
the finite-time future singularities appear have been
reconstructed (a similar investigation has been executed also in Ref.~\cite{Setare:2012vs}).

In this paper, we study the features of
dark energy cosmologies in the context of LQC.
It is shown that in conventional cosmology,
at the dark energy dominated stage,
the domination of dark energy
finite-time or infinite-time (such as the Little Rip~\cite{fls11,
Brevik:2011mm, Frampton:2011rh} and Pseudo-Rip~\cite{fls12} scenarios)
future singularities may occur,
but that in LQC, some of these singularities may be avoided.

In this theory, holonomy corrections lead to $\rho^2$ correction term
in the Friedmann equation at high energies~\cite{s06}, which give rise
to a bounce when the energy density becomes equal to a critical value of the
order of the Planck energy density.
Thus, this bounce prevents the existence of Rip singularities in LQC.
Indeed, the Friedmann equation in LQC is an ellipse in the plane $(H,\rho)$.
This means that the universe moves along this ellipse and the values of
the variables $H$ and $\rho$
are constrained to be bounded~\cite{h12a, h12b, h12c}.
This is the main difference from the FLRW cosmology, where
the Friedmann equation describes a parabola in the plane $(H,\rho)$.
Accordingly, in the FLRW cosmology,
the universe moves along this parabola and
the variables $H$ and $\rho$ can take unbounded values.
Hence, for dark energy models described with those EoS,
Rip singularities could occur.
On the other hand, singularities that appear for finite values of the energy
density and the Hubble parameter, like sudden singularities,
could still happen in LQC.

Furthermore, we generalize our analysis on future singularities
to $F(T)$ gravity.
For this procedure,
LQC is one of the best examples.
We show that in this theory,
the universe moves along a curve in the plane $(H,\rho)$,
given by the corresponding modified Friedmann equation,
and that the dynamics is described by the conservation and the corresponding
modified Raychauduri equations.
Finally, we present several non trivial examples of the EoS
where future singularities appear.
We use units of $k_\mathrm{B} = c = \hbar = 1$ and denote the
gravitational constant $8 \pi G_\mathrm{N}$ by
${\kappa}^2 \equiv 8\pi/{M_{\mathrm{Pl}}}^2 = 1$
with the Planck mass of $M_{\mathrm{Pl}} = G_\mathrm{N}^{-1/2} = 1.2 \times
10^{19}$GeV.

The paper is organized as follows.
In Sec.~II, we review Einstein cosmology (EC) and the singularities appearing
in it.
In Sec.~III, we consider LQC and derive the modified Friedmann and Raychauduri
equation.
In Sec.~IV,
we study future singularities in LQC and show that the Rip singularities do not
survive, but Type II and Type IV singularities could still happen.
Section V is devoted to investigate Pseudo-Rip models, where the universe
asymptotically approaches to the de Sitter regime. We illustrate that
disintegration of bound structures could occur in LQC, provided that
it previously occurs in EC.
In Sec.~VI, we investigate teleparallel cosmological theories and
demonstrate that LQC and Brane cosmology in the Randall-Sundrum
scenario belong to this category.
Finally, conclusions are presented in Sec.~VII.

\section{Einstein cosmology}

For the flat FLRW space-time with the metric
$ds^2 = dt^2 - a^2(t) d{\Vec{x}}^2$,
where $a$ is the scale factor,
filled by a perfect fluid with the
state equation $P=-\rho-f(\rho)$, EC is obtained from the Lagrangian
${\mathcal L}=\frac{1}{2}R V-\rho V$ where $R=6(\dot{H}+2H^2)$ is the scalar
curvature, $V=a^3$ is the volume and $H=\frac{\dot{a}}{a}$
is the Hubble parameter.
Here, the dot denotes the time derivative of $\partial/\partial t$.

This Lagrangian has been constructed in co-moving fluid coordinates
(see \cite{s03} and Sec.~III C of~\cite{r72}), and
the energy density $\rho$ has to be understood as a function of the volume $V$.
This relation comes from the
conservation equation
\begin{eqnarray}\label{1}
  d(\rho V)=-Pd(V)\Longleftrightarrow
\dot{\rho}=-3H(\rho+P)\Longleftrightarrow \frac{d\rho}{dV}
=-\frac{1}{V}(\rho+P)\,.
\end{eqnarray}

{}From $P=-\rho-f(\rho)$, we find the differential equation
\begin{eqnarray}\label{2}
  \frac{d\rho}{f(\rho)}=\frac{dV}{V}\,.
\end{eqnarray}
After integration, we obtain the expression of $\rho$ as a function of $V$. For
example, if $f(\rho)=-(w+1) \rho$, one has
\begin{eqnarray}\label{3}
  \rho(V)=\rho_0\left(\frac{V}{V_0}\right)^{-(1+w)}\,,
\end{eqnarray}
where $\rho_0$ is the value of $\rho$ for $V=V_0$.

Note that the Lagrangian can be written as follows:
${\mathcal L}=\left(3\frac{d\dot{V}}{dt}-\frac{1}{3 V}\dot{V}^2 \right)-\rho
V$. This means that the same theory
is acquired by removing the total derivative,
and eventually we have
the Lagrangian ${\mathcal L}_{E}=-\frac{1}{3V}\dot{V}^2  -\rho V$.
The conjugate momentum is given by
$p_V=\frac{\partial {\mathcal L}_{E}}{\partial\dot{V}}= -\frac{2}{3V}\dot{V}
=-2H$, and thus the Hamiltonian becomes
\begin{eqnarray}\label{4}
{\mathcal H}_{E}=
\dot{V}p_V- {\mathcal L}_{E}= -\frac{1}{3V}\dot{V}^2  +\rho V
=-\frac{3}{4}p_V^2V  +\rho V=
-3H^2 V+\rho V\,.
\end{eqnarray}

In general relativity, the Hamiltonian is constrained to be zero.
This leads to the Friedmann equation
\begin{eqnarray}\label{5}
H^2=\frac{\rho}{3}\,.
\end{eqnarray}
This equation represents a parabola in the plane $(H,\rho)$, that is, the
evolution of the universe follows this parabola, and its dynamics is given by
the so-called Raychauduri and conservation equations.
The Raychauduri equation is obtained from the Hamilton equation
$\dot{p}_V=-\frac{\partial {\mathcal H}_E}{\partial V}$, which
reads
\begin{eqnarray}\label{6}
\dot{p}_V= \frac{3}{4}p_V^2-\frac{\partial (\rho V)}{\partial V}= \rho+P\,,
\end{eqnarray}
where we have used the conservation and Friedmann equations.
Since $p_V=-2H$, we eventually obtain
the Raychauduri equation $\dot{H}=-\frac{1}{2}(\rho+P)=\frac{f(\rho)}{2}$.
Hence, the dynamics of the universe is given by
\begin{eqnarray}\label{7}\left\{\begin{array}{ccc}
  \dot{H} \Eqn{=} \frac{f(\rho)}{2}\\
\dot{\rho} \Eqn{=} 3Hf(\rho),\end{array}\right.
\end{eqnarray}
provided that the universe moves along the parabola $\rho=3H^2$.

\begin{remark}
Since $\rho=3H^2$, the dynamics of the system is given by the following
$1$-dimensional first-order differential equations
\begin{eqnarray}\label{8}
  \dot{H} = \frac{f(3H^2)}{2}.
\end{eqnarray}
\end{remark}

For the system (\ref{7}), future singularities are bound to appear in a finite
time $t_s$. Actually,
they can be classified as follows (for details, see~\cite{not05, f10}):
\begin{enumerate}
\item Type I (Big Rip): For $t\rightarrow t_s$, $a\rightarrow \infty$,
$\rho\rightarrow \infty$ and
$|P|\rightarrow \infty$.

\item Type II (Sudden): For $t\rightarrow t_s$, $a\rightarrow a_s$,
$\rho\rightarrow \rho_s$ and
$|P|\rightarrow \infty$.

\item Type III (Big Freeze): For $t\rightarrow t_s$, $a\rightarrow a_s$,
$\rho\rightarrow \infty$ and
$|P|\rightarrow \infty$.

\item Type IV (Generalized Sudden): For $t\rightarrow t_s$, $a\rightarrow a_s$,
$\rho\rightarrow 0$ and
$|P|\rightarrow 0$ and higher derivatives of $H$ diverge.
The case that $\rho$ and/or $|P|$
tend to finite values in the limit $t \rightarrow t_s$
is also included~\cite{Shtanov:2002ek}.
\end{enumerate}

There also exist future singularities at infinite time
like the Little Rip (LR)~\cite{fls11},
defined as follows:
$$H(t)\rightarrow \infty, \quad \mbox{when}\quad t\rightarrow \infty.$$
The condition that this singularity exists could
easily be deduced from Eq.~(\ref{8}) and it reads
\begin{eqnarray}\label{9}
  \int_{H_0}^{\infty}\frac{d H}{f(3H^2)}=\infty\,.
\end{eqnarray}
As an example,
for the case $f(\rho)=A\rho^{1/2}$, we find~\cite{fls11}
\begin{eqnarray}\label{10}
  H(t)=H_0e^{\frac{\sqrt{3}}{2}A(t-t_0)}\,,
\end{eqnarray}
where $H_0=H(t_0)$.

\section{Loop quantum cosmology (LQC)}

An approach to quantum cosmology that could avoid singularities is LQC.
The main idea is that LQC assumes a discrete nature of space which leads,
at quantum level, to consider a Hilbert space where
quantum states are represented by almost periodic functions of the dynamical
part of the connection~\cite{abl03}.
Unfortunately, the connection variable does not correspond to a well defined
quantum operator in this Hilbert space and therefore we
  re-express the gravitational part of the Hamiltonian in terms of almost
periodic function. It could be executed from a process of regularization.
This new regularized Hamiltonian introduces a quadratic modification
($\rho^2$) in the Friedmann equation at high energies~\cite{s06, svv06},
which give rise to a bounce when the energy density becomes equal to a critical
value of the order of the Planck energy density.

This theory is constructed as follows.
The old quantization of LQC was done by using two canonically conjugate
variables. One is the dynamical part of the connection, i.e., ${\mathfrak c}$,
and
the other is the dynamical part of the triad, namely, ${\mathfrak p}$.
These variables are related with the scale factor and
the extrinsic curvature $K=\frac{1}{2}\dot{a}$ by the relations
(for instance, see~\cite{abl03, a07})
\begin{eqnarray}\label{a1}
  |{\mathfrak p}|=a^2, \qquad {\mathfrak c}=2\gamma K={\gamma}\dot{a},
\end{eqnarray}
where  $\gamma\cong 0.2375$ is the Barbero-Immirzi parameter \cite{m04},
and their Poisson bracket is given by $\{{\mathfrak c},{\mathfrak
p}\}=\frac{\gamma}{3}\mbox{sgn}({\mathfrak p})$.

To build the quantum theory in LQC,
as a Hilbert space, we usually choose
the quotient space as the Besicovitch space of
almost periodic functions by its subspace of null functions.
The Besicovitch space (for details, see~\cite{b54})
is the closure of trigonometric polynomials under the
semi-norm (in the ${\mathfrak c}$-representation)
$$||\Psi||^2=\lim_{L\rightarrow \infty}\frac{1}{2L}\int_{-L}^{L}
|\Psi({\mathfrak c})|^2 d{\mathfrak c}\,,
$$ where ${\mathfrak c}$ is the connection.
Also, all the element of this space have the expansion
$$\Psi({\mathfrak c})=\sum_{n\in\Z}\alpha_n
|\mu_n\rangle\equiv\sum_{n\in\Z}\alpha_ne^{i\mu_n{\mathfrak c}/2},$$
with $\mu_n\in\R$ and $\alpha_n\in {\mathfrak l}^2$ (the space of
square-summable sequences).

In this space, we can define the operator $\hat{{\mathfrak p}}$ as
$\hat{{\mathfrak p}} \equiv -\frac{i\gamma}{3}\frac{d}{d{\mathfrak c}}$.
However, the operator $\hat{\mathfrak c}$ defined by $\hat{\mathfrak
c}\Psi({\mathfrak c}) \equiv {\mathfrak c}\Psi({\mathfrak c})$ is not well
defined in this Hilbert space because for a general quantum state
$\Psi({\mathfrak c})=\sum_{n\in\Z}\alpha_ne^{i\mu_n{\mathfrak c}/2}$,
we obtain
$$||\hat{\mathfrak c}\Psi||^2=\lim_{L\rightarrow
\infty}\frac{L^2}{3}\sum_{n\in\Z}|\alpha_n|^2=+\infty\,.$$


It is important to stress a key point in LQC.
The gravitational part of the Hamiltonian in EC contains
${\mathfrak c}$. In fact, we have
$${\mathcal H}_{grav, EC}=-3H^2V= -\frac{3}{\gamma^2}{\mathfrak c}^2|{\mathfrak
p}|^{1/2}\,.$$
Since the operator $\hat{\mathfrak c}$ is not well-defined,
in order to build the quantum theory, we need to 
re-define the gravitational part of the Hamiltonian. To do that,
we introduce the new canonically conjugate 
variables~\cite{as11,s09a}:
\begin{eqnarray}\label{variables}
  \beta\equiv \frac{{\mathfrak c}}{|{\mathfrak p}|^{1/2}}=\gamma H, \qquad
|{\mathcal V}|=|{\mathfrak p}|^{3/2}=V,
\end{eqnarray}
with Poisson bracket $\{\beta,{\mathcal
V}\}=\frac{\gamma}{2}\mbox{sgn}({\mathcal V})$.
We work in the $\beta$ representation 
and consider the holonomies 
$h_j(\lambda)\equiv \e^{-i\frac{\lambda \beta}{2}\sigma_j}=\cos(\frac{\lambda\beta}{2})-i\sigma_j\sin(\frac{\lambda\beta}{2})$ (see for details Section $2$ of~\cite{s09a}, Sections II D and II E of~\cite{as11}, or~\cite{he10}), where 
$\sigma_j$ are the Pauli matrices 
and $\lambda$ is a parameter with the dimension of length, which numerical value is determined by invoking the quantum nature of the geometry, that is, 
identifying its square with the minimum eigenvalue of the area operator in 
LQG. As a consequence, it follows $\lambda \equiv 
\sqrt{\frac{\sqrt{3}}{4}\gamma}$ (see~\cite{s09a}).

Thus, using the variables (\ref{variables}) one has ${\mathcal H}_{grav, EC}= -\frac{3}{\gamma^2}\beta^2V,$
and since
$\beta^2$, for the same reason as $\mathfrak{c}^2$, does not have a well-defined quantum operator,
in order to construct a consistent quantum Hamiltonian operator,
we need an almost periodic function that approaches $\beta^2$ for small
values of $ \beta$.
This can be executed
with the general formulae of LQC~\cite{t01, cm10, aps06} to
acquire the regularized Hamiltonian~\cite{he10, dmw09}
\begin{eqnarray}\label{a2}
H_{grav,LQC} \Eqn{\equiv} -\frac{2}{\gamma^3\mu^3}
\sum_{i,j,k}\varepsilon^{ijk}Tr\left[
h_i(\lambda)h_j(\lambda)h_i^{-1} (\lambda)
h_j^{-1}(\lambda)h_k(\lambda)\{h_k^{-1}(\lambda),V\}\right]\nonumber\\
\Eqn{=} -\frac{3V}{\gamma^2\lambda^2}\sin^2(\lambda \beta)\,,
\end{eqnarray}
which captures the underlying loop quantum dynamics (see also ~\cite{abl03, aps06,s09, s09a}).
Note that this Hamiltonian admits a quantum version because the operators
$\hat{h}_j(\lambda)$ are well-defined in this Hilbert space.


Then the total
 effective Hamiltonian is
\begin{eqnarray}\label{11}
{\mathcal H}_{LQC}=-3V\frac{\sin^2( \lambda \beta)}{\gamma^2\lambda^2}+
V,
\end{eqnarray}
and
the Hamiltonian constraint is given by
$\frac{\sin^2( \lambda\beta)}{\gamma^2\lambda^2}=\frac{\rho}{3}$.
The Hamiltonian equation gives the following identity:
\begin{eqnarray}\label{13}
\dot{V}=\{V,{\mathcal H}_{LQC}\}=-\frac{\gamma}{2}\frac{\partial{\mathcal
H}_{LQC}}{\partial\beta}
\Longrightarrow
H= \frac{\sin(2\lambda \beta)}{2\gamma\lambda}\Longleftrightarrow \beta=
\frac{1}{2\lambda}\arcsin(2\lambda\gamma H).
\end{eqnarray}
By writing this last equation as $H^2=\frac{\sin^2(\lambda
\beta)}{\gamma^2\lambda^2}(1-\sin^2( \lambda \beta))$ and
using the Hamiltonian constraint ${\mathcal H}_{LQC}=0\Longleftrightarrow
\frac{\sin^2( \lambda \beta)}{\gamma^2\lambda^2}=\frac{\rho}{3}$,
we find
the following modified Friedmann equation in LQC
\begin{eqnarray}\label{14}
H^2=\frac{\rho}{3}\left(1-\frac{\rho}{\rho_c}\right)
\Longleftrightarrow
\frac{H^2}{\rho_c/12}+\frac{(\rho-\frac{\rho_c}{2})^2}{\rho_c^2/4}=1\,,
\end{eqnarray}
with $\rho_c\equiv \frac{3}{\gamma^2\lambda^2}$.
This modified Friedmann equation is an ellipse in the plane $(H,\rho)$
(a parabola in EC), and the dynamics is given by
the system
\begin{eqnarray}\label{15}\left\{\begin{array}{ccc}
\dot{H} \Eqn{=} \frac{f(\rho)}{2}\left(1-\frac{2\rho}{\rho_c} \right)\\
\dot{\rho} \Eqn{=} 3Hf(\rho)\,,\end{array}\right.
\end{eqnarray}
where the first equation is the modified Rauchauduri equation in LQC.


\begin{remark}
The modified Friedmann equation in LQC (i.e., Eq.~(\ref{14})) appears
in brane cosmology (BC) in the Randall-Sundrum scenario~\cite{aeoy12, s06},
where the modified Friedmann equation reads
\begin{eqnarray}\label{16}
H^2=\frac{\rho}{3}\left(1+\frac{\rho}{2\Lambda}\right)\,,
\end{eqnarray}
with $\Lambda$ the brane tension.
Hence, if we take $\Lambda=-\rho_c/2$, Eq.~(\ref{16}) becomes Eq.~(\ref{14}),
that is, BC with negative brane tension is equivalent to LQC.
On the other hand, the modified Friedmann equation in BC with positive brane
tension could be deduced from the Hamiltonian
${\mathcal H}_{BC}=-3V\frac{\sinh^2( \lambda \beta)}{\gamma^2\lambda^2}+
V\rho$. In the same way, we have obtained Eq.~(\ref{14}). For the
Hamiltonian ${\mathcal H}_{BC}$, we find
$$H^2=\frac{\rho}{3}\left(1+\frac{\rho}{2\Lambda}\right)\,,
$$
where now $\Lambda=\rho_c/2$.
\end{remark}

Finally, we remark that for the following parameterization of the ellipse
\begin{eqnarray}\label{17}
H=\sqrt{\frac{\rho_c}{12}}\cos(\eta),\quad
\rho=\frac{\rho_c}{2}(1+\sin\eta),\quad 0\leq \eta\leq 2\pi\,,
\end{eqnarray}
%
the dynamics in the case of (\ref{15}) is now given by 
%
\begin{eqnarray}\label{18}
\dot{\eta}=\sqrt{\frac{3}{\rho_c}}F(\eta)\,,
\end{eqnarray}
where we have introduced the notation 
$F(\eta)=f(\frac{\rho_c}{2}(1+\sin\eta))$. This means that in LQC with a state
equation of the form
$P=-\rho-f(\rho)$, we analyze a $1$-dimensional first order differential
equation.

\section{Future singularities in LQC}

Future singularities have been studied
%
by using the effective approach of LQC
%
in Refs.~\cite{sst06,nw07,sg07,s09,lw10,cs09,sv11,ccvw08}.
In this section, we examine what kind of future singularities could appear
%
with the effective approach of LQC
%
by following our own novel view point.

\subsection{Rip singularities}

Rip singularities such as the Big Rip and Little Rip
are characterized by the fact that
$H\rightarrow \infty$ and/or $\rho\rightarrow \infty$.
However, in LQC, from the equation of the ellipse in (\ref{14}), we can deduce
that $\rho$ is constrained to belong to the interval $[0,\rho_c]$ and $H$ in
the interval $[-\sqrt{\frac{\rho_c}{12}}, \sqrt{\frac{\rho_c}{12}}]$.
This means that these singularities cannot appear in LQC.
For the same reason, the Big Freeze singularity does not occur neither.
This behavior is very different from those in EC and BC with positive brane
tension. Indeed, in EC, the Friedmann equation is
a parabola $\rho=3H^2$ in the plane $(H,\rho)$,
and in BC with positive brane tension (Eq.~(\ref{16}) with $\Lambda>0$),
the Friedmann equation is the hyperbola
$\frac{(\rho-\Lambda)^2}{\Lambda^2}-\frac{H^2}{\Lambda/6}=1
$.
It is significant that in both cases, the variables $H$ and $\rho$
are allowed to diverge, and hence for a model with the certain EoS,
the Rip singularity appears.
In order to clarify this point, we present the helpful two examples.

\begin{enumerate}
  \item
As a first example, we consider the model $f(\rho)=\alpha^2$ with $\alpha$ a
constant~\cite{aeoy12}.
In EC, after integration of the system (\ref{7}), we obtain
\begin{eqnarray}\label{19}
  H(t)=\frac{\alpha^2}{2}(t-t_0)+H_0,\quad
\rho(t)=\frac{3\alpha^4}{4}(t-t_0)^2+3\alpha^2H_0(t-t_0)+3H_0^2\,,
\end{eqnarray}
which gives a LR singularity.
In BC with positive brane tension ($\Lambda>0$), the conservation equation
reads
\begin{eqnarray}\label{20}
  \dot{\rho}={3\alpha^2}\sqrt{\frac{\Lambda}{3}}\sqrt{\frac{(\rho-\Lambda)^2}{\Lambda^2}-1}\,,
\end{eqnarray}
which could be integrated and accordingly giving the following result:
\begin{eqnarray}\label{21}
  {\rho}(t)=\Lambda \left[1+
\cosh\left({3\alpha^2}\sqrt{\frac{1}{6\Lambda}}(t-t_0)+\cosh^{-1}\left(\frac{\rho_0-\Lambda}{\Lambda}\right)\right)\right].
\end{eqnarray}
When $t\rightarrow \infty$, we have a LR with the following dynamics:
\begin{eqnarray}\label{22}
  H(t)\sim \frac{\Lambda}{6} e^{{3\alpha^2}\sqrt{\frac{1}{6\Lambda}}t}
,\quad{\rho}(t)
\sim\Lambda e^{{3\alpha^2}\sqrt{\frac{1}{6\Lambda}}t}\,.
\end{eqnarray}
However, if the brane tension is negative (i.e., in LQC), we acquire
\begin{eqnarray}\label{23}
  {\rho}(t)=\Lambda \left[1+
\sin\left({3\alpha^2}\sqrt{\frac{1}{6\Lambda}}(t-t_0)+\sin^{-1}\left(\frac{\rho_0-\Lambda}{\Lambda}\right)\right)\right]\,,
\end{eqnarray}
that is, $\rho$ always remains in the interval $[0,2\Lambda]$.

\item
Another example is given by the model $f(\rho)=A\rho^{1/2}$ with
$A$ a constant~\cite{fls11}.
As seen in Eq.~(\ref{10}), in EC, we have the solution
\begin{eqnarray}\label{24} H(t)=H_0e^{\frac{\sqrt{3}}{2}A(t-t_0)}, \quad
  {\rho}(t)=3H_0^2e^{\sqrt{3}A(t-t_0)}\,,
\end{eqnarray}
which give rise to a LR singularity.
On the other hand, if we investigate the modified Friedmann in LQC (namely,
Eq.~(\ref{14})) and substitute $H$ into the conservation equation,
in the expanding phase ($H>0$) we acquire
$\dot{\rho}=\sqrt{3}\rho\sqrt{1-\frac{\rho}{\rho_c}}$.
This equation could be integrated and eventually we find
\begin{eqnarray}\label{25}
  \rho(t)=\rho_c\left[1-\left(\frac{1-e^{\sqrt{3}|t-t_c|}}{1+e^{\sqrt{3}|t-t_c|}}\right)^2\right]\,,
\quad
-\infty< t <+\infty\,,
\end{eqnarray}
where $t_c$ is the time when the universe arrives at the point $(0,\rho_c)$.
This solution shows that a universe moves in an anti-clockwise sense,
and bounces (namely, it enters in the contracting phase) at time $t_c$.
This means that there does not happen the LR in LQC.
\end{enumerate}

\subsection{Type II and Type IV singularities}

When $f(\rho)$ diverges at some point $\rho_s$, sudden singularities appear
because at this point $P(\rho_s)=\infty$.
Clearly,
this singularity occurs in EC when $0\leq \rho_s<\infty$, and it appears in LQC
when $0\leq \rho_s\leq\rho_c$.
In order to compare sudden singularities
in EC with those in LQC, we assume that $0\leq \rho_s\leq\rho_c$.
Note that the time when the system arrives at the singularity is different
between  EC and LQC. Effectively, from Eqs.~(\ref{7}) and (\ref{15}),
we deduce
\begin{eqnarray}\label{26}
  t_{s,EC}=t_0+\int_{\rho_0}^{\rho_s}\frac{d\rho}{\sqrt{3\rho}f(\rho)},\quad
t_{s,LQC}=t_0+\int_{\rho_0}^{\rho_s}\frac{d\rho}{\sqrt{3\rho}
\sqrt{1-\frac{\rho}{\rho_c}}
f(\rho)}.
\end{eqnarray}
Equation (\ref{26}) shows that for phantom dark energy ($f(\rho)\geq 0$),
we always have $t_{s,EC}\leq t_{s,LQC}$
because $\sqrt{1-\frac{\rho}{\rho_c}}\leq 1$.

An interesting quantity is the inertial force $F_{inert}$ on a mass $m$
measured by an observer at a comoving distance $l$, and it is given by
\begin{eqnarray}\label{27}
  F_{inert}=ml\frac{\ddot{a}}{a}=ml(\dot{H}+H^2)\,.
\end{eqnarray}
Near the sudden singularity, since $\dot{H}$ diverges and $H$ converges,
we find
\begin{eqnarray}\label{28}
  F_{inert}\cong ml\dot{H}\,.
\end{eqnarray}
Hence, near the sudden singularity, we obtain
the following expression in EC and LQC
\begin{eqnarray}\label{29}
  F_{inert,EC}\cong ml\frac{f(\rho)}{2}, \quad  F_{inert,LQC}\cong
ml\frac{f(\rho)}{2}\left(1-\frac{2\rho}{\rho_c}\right)\,.
\end{eqnarray}
{}From this last equation, we deduce that for a phantom dark model,
$F_{inert,EC}$ is always positive. However, $F_{inert,LQC}$ is positive
for $0\leq\rho\leq \rho_c/2$, whereas it is negative for $\rho_c/2\leq \rho\leq
\rho_c$. We can also conclude that $|F_{inert,LQC}|\leq F_{inert,EC}$, i.e.,
near the Type II singularity the inertial force in EC is always greater than
that in LQC. Note that the inertial force is important because when it exceeds
the (gravitational) force bounding the system then disintegration of bound
objects occurs.

In summary,
%
as shown numerically in early works,
%
Type II and Type IV singularities could survive in LQC.
Indeed, they survive provided $0\leq \rho_s\leq\rho_c$.
On the other hand, when
%
$\rho_c<\rho_s$,
%
the singularity disappears and
the universe becomes cyclic moving in an anti-clockwise sense
along the ellipse.

Type IV singularities always survive in LQC.
This is because they appear at the point $(0,0)$ and near this point
the dynamical systems (\ref{7}) and (\ref{15}) coincide with each other.
As an example, we consider the case $f(\rho)=-\alpha^2\rho^{\gamma}$ with
$0<\gamma<1/4$. Near $(0,0)$, the conservation equation reads
$\dot{\rho}\sim -\sqrt{3}\alpha^2\rho^{\frac{2\gamma+1}{2}}$, whose solution
is given by
\begin{eqnarray}\label{30}
\rho(t)\sim \left(-\frac{\sqrt{3}(1-2\gamma)\alpha^2}{2}(t-t_0)+
\rho_0^{\frac{1-2\gamma}{2}}\right)^{\frac{2}{1-2\gamma}}\,.
\end{eqnarray}
This means that the system arrives at the point $(0,0)$
at $t = t_s\sim
\frac{2}{\sqrt{3}(1-2\gamma)\alpha^2}\rho_0^{\frac{1-2\gamma}{2}}+t_0$, where
all the derivatives of higher order than one
of $H$ diverge.

\section{Pseudo-Rip models}

The Pseudo-Rip (PR) has been proposed in Ref.~\cite{fls12}. It is defined by
$P \rightarrow -\rho$ as $t\rightarrow \infty$, provided $f(\rho)\geq 0$.
In this case, the (mild phantom) universe asymptotically approaches to the de
Sitter regime.

As a first example of PR, we consider the model $H(t)=H_0-H_1e^{-\gamma t}$
with $\gamma>0$~\cite{Frampton:2011rh}. To obtain the EoS
corresponding to this example in EC, by inserting this expression
into the Friedmann and Raychauduri equations and
eliminating the variable $t$, we get
\begin{eqnarray}\label{31}
  f(\rho)=2H_0\gamma\left(1-\sqrt{\frac{\rho}{3H_0^2}}\right)\,,\quad \mbox{
for}\quad 0\leq \rho\leq 3H_0^2\,.
\end{eqnarray}

On the other hand, the corresponding EoS in LQC is obtained in the same way
by using the corresponding modified equations.
The final result becomes
\begin{eqnarray}\label{32}
  f(\rho)=\frac{2H_0\gamma}{1-\frac{2\rho}{\rho_c}}\left((1-\sqrt{\frac{\rho}{3H_0^2}\left(1-\frac{\rho}{\rho_c}\right)}\right)\,,
\quad \mbox{ for}\quad 0\leq \rho\leq
\frac{\rho_c}{2}\left(1-\sqrt{1-\frac{12H_0^2}{\rho_c}}\right)\,,
\end{eqnarray}
provided $12H_0^2<\rho_c$.

When PR appears, the quantity of interest is the inertial force
$F_{inert}$. In EC, it is given by
\begin{eqnarray}\label{33}
  F_{inert, EC}=ml\left(\frac{f(\rho)}{2}+\frac{\rho}{3}\right)\,,
\end{eqnarray}
and in LQC it is described as
\begin{eqnarray}\label{34}
  F_{inert,LQC}=ml\left(\frac{f(\rho)}{2}\left(1-\frac{2\rho}{\rho_c}\right)+\frac{\rho}{3}\left(1-\frac{\rho}{\rho_c}\right)\right)\,.
\end{eqnarray}
Since $f(\rho)\geq 0$ (the phantom model),
we see that $|F_{inert,LQC}|\leq F_{inert,EC}$,
and thus the disintegration of bound structures
could occur in LQC, only if previously it occurs in EC.
More precisely,
let $F_{\Omega}$ be the force which keeps bounded the structures.
Since structures disintegrate when $F_{inert}\geq F_{\Omega}$ and when
$\rho\longrightarrow \rho_f$, we find
\begin{eqnarray}\label{35}
  F_{inert, EC}\cong ml\frac{\rho_f}{3}\,, \quad
  F_{inert,LQC}\cong ml\frac{\rho_f}{3}\left(1-\frac{\rho_f}{\rho_c}\right)\,.
\end{eqnarray}
Accordingly, we can conclude that in EC, structures disintegrate if
\begin{eqnarray}\label{36}
  F_{\Omega}\leq \frac{ml\rho_f}{3}\,.
\end{eqnarray}

On the other hand,
%
using the expression is (\ref{14}),
%
in LQC, structures disintegrate for
\begin{eqnarray}\label{37}
  \frac{\rho_c}{2}\left(1-\sqrt{1-\frac{12 F_{\Omega}}{ml\rho_c} } \right)\leq
\rho_f\leq
\frac{\rho_c}{2}\left(1+\sqrt{1-\frac{12 F_{\Omega}}{ml\rho_c} } \right)\,,
\end{eqnarray}
provided
\begin{eqnarray}\label{38}
  F_{\Omega}\leq \frac{ml\rho_c}{12}\,.
\end{eqnarray}
Note that if the condition (\ref{38}) is satisfied, then
the condition (\ref{36}) will also be met automatically, because
$\rho_f\leq \rho_c$.
This proves our statement that disintegration of bound structures
could happen in LQC, only if previously it occurs in EC.
In other words, the disintegration time in LQC is always bigger, the
corresponding universe is effectively more stable than that in EC.

\begin{remark}
In BC with positive brane tension $\Lambda$, near the PR we get
\begin{eqnarray}\label{39}
  F_{inert,BC}\cong ml\frac{\rho_f}{3}\left(1+\frac{\rho_f}{2\Lambda}\right)\geq
F_{inert, EC}\,,
\end{eqnarray}
that is, if disintegration of bound structures
occur in EC, then they also happen in BC with positive brane tension.
\end{remark}

\section{Teleparallel cosmological theories}

LQC is characterized by two important items:
\begin{enumerate}
\item
The canonically conjugate variables $(V,\beta)$ are no longer defined by
$V=a^3$ and $\beta=\gamma H$. In LQC, from Eq.~(\ref{13}) we conclude that
they are defined by $V=a^3$ and
$\beta=\frac{1}{2\lambda}\arcsin(2\lambda\gamma H)$, and that
only when $\lambda\rightarrow 0$, we take the early form.

\item

{}From the relation $\{\beta,V\}=\frac{\gamma}{2}\Longleftrightarrow \{V,-\frac{2}{\gamma}\beta\}=1$, and following the Hamiltonian formulation
one deduces that $V$ plays the role of the position
and $-\frac{2}{\gamma}\beta$ is its corresponding congugate momentum, then
 Legendre's transformation gives us
\begin{eqnarray}\label{40}
{\mathcal H}_{LQC}=-\frac{2}{\gamma}\dot{V}\beta-{\mathcal L}_{LQC}\,,
\end{eqnarray}
we obtain the Lagrangian in LQC.

In terms of the variables $V$ and $H$, it reads
\begin{eqnarray}\label{41}
  {\mathcal L}_{LQC}=-\frac{3VH}{\gamma\lambda}\arcsin(2\lambda\gamma H)+
\frac{3V}{\gamma^2\lambda^2}\sin^2\left(\frac{1}{2}\arcsin(2\lambda\gamma H)
\right)-V\rho\,,
\end{eqnarray}
which coincides with ${\mathcal L}_{E}$ for small values of $\lambda$.
Note that this Lagrangian does not belong to the category
of modified gravity models. It is not built from the
scalar curvature $R=6\left(\dot{H}+2H^2\right)$
and the Gauss-Bonnet curvature invariant
$G=24H^2\left(\dot{H}+H^2\right)$, i.e., it does not have the form
${\mathcal L}_{GB}=VF(R,G)-V\rho$ for some function $F$.
Insteatd of it, LQC belongs to the category of the so-called
{\it teleparallel theories} of General Relativity~\cite{hs79, g10}.
\end{enumerate}

\subsection{Teleparallelism}

Teleparallel theories are based in the Weitzenb\"ock space-time.
To build this space-time, we choose a global system of four orthonormal
vector fields $\{{\bf e}_i\}$ related to the vectors
$\{\partial_{\mu}\}$ via the relation ${\bf e}_i=e_i^{\mu}\partial_{\mu}$.
In addition, we introduce a covariant derivative $\nabla$ that defines
absolute parallelism with respect the global basis $\{{\bf e}_i\}$,
that is, $\nabla_{\nu}e_i^{\mu}=0$.
{}From this, we acquire the metric
Weitzenb\"ock connection $\Gamma^{\gamma}_{\quad\mu
\nu}=e_i^{\gamma}\partial_{\nu}e^i_{\mu}$. (Note that this connection is metric
and therefore it
satisfies $\nabla_{\gamma}g_{\mu\nu}=0$.)

The Weitzenb\"ock space-time has identically vanishing curvature (the Riemann
tensor vanishes), but it is not torsion free.
Effectively, we have
${T^{\gamma}}_{\mu \nu}={\Gamma^{\gamma}}_{\nu\mu}-{\Gamma^{\gamma}}_{\mu \nu}
\neq 0$.
In order to describe the Lagrangian in the Weitzenb\"ock space-time,
we introduce the contorsion tensor
$${K^{\mu\nu}}_{\gamma}=-\frac{1}{2}\left({T^{\mu\nu}}_{\gamma}
- {T^{\nu\mu}}_{\gamma} - {T^{\mu\nu}}_{\gamma}
\right)\,,$$
and the tensor
$$ {S^{\mu\nu}}_{\gamma}=\frac{1}{2}\left( {K^{\mu\nu}}_{\gamma} +
\delta^{\mu}_{\gamma} {T^{\theta \nu}}_{\theta} -
\delta^{\nu}_{\gamma} {T^{\theta \mu}}_{\theta} \right)\,.$$
Then, we can construct the invariant
$$T= {S^{\mu\nu}}_{\gamma} {T^{\gamma}}_{\mu \nu}\,,$$
and define modified teleparallel theories via the Lagrangian of the
form ${\mathcal L}_{T}=VF(T)-V\rho$.

The interesting point is that if we take the basis $\{{\bf e}_0=\partial_{0},
{\bf e}_1=a\partial_{1}, {\bf e}_2=a\partial_{2},
  {\bf e}_3=a\partial_{3}\}$, then for the FLRW metric
we find~\cite{bf09, bmno12} $T=-6H^2$, and hence we acquire
${\mathcal L}_{LQC}=VF(T)-V\rho$ with the choice
\begin{eqnarray}\label{42}
F(T)= -\frac{3\sqrt{-T/6}}{\gamma\lambda}\arcsin(2\lambda\gamma \sqrt{-T/6})+
\frac{3}{\gamma^2\lambda^2}\sin^2\left(\frac{1}{2}\arcsin(2\lambda\gamma
\sqrt{-T/6})
\right)\,.
\end{eqnarray}

The problematic point in teleparallelism is that
it depends on the choice of the global basis.
In the sense that, if one uses a non-local
Lorentz transformation to transform the original  global basis in another one,
in general, one obtains another Weitzenb\"ock connection and thus $T$ could
change~\cite{lsb11, Li:2011rn}, which does not happen in
modified gravity where the invariants do not depend of a global basis.
%
%
However, for a FLRW metric, if we use
local Lorentz transformations that only depend on the time, that is,
of the form ${\bf \bar{e}}_i=\Lambda^k_i(t){\bf e}_k$,
then, even though the torsion tensor changes, the torsion scalar $T$ remains
constant with a value of $T=-6H^2$.
{}From our point of view, this gives a consistency
with teleparallel theories in
cosmology (and in particular, with LQC), because
$T$ is invariant from ``isotropic and homogeneous''
local Lorentz transformations.

\subsection{Singularities in teleparallel theories}

Singularities in teleparallel theories have recently been explored
in great detail in Ref.~\cite{bmno12}.
Here, we present several remarkable results.
First,
in modified teleparallel theories the modified Friedmann equation gives a
curve in the plane $(H,\rho)$
(as we have already seen, in LQC, it is an ellipse, and
in EC, it is a parabola, whereas in BC with positive brane tension,
it is an hyperbola). Thus,
if this curve is a bounded set, the Rip and Freeze singularities do not
survive.

Now, we consider teleparallel theories with a Lagrangian on the form
${\mathcal L}_{T}=VF(T)-V\rho$. In these theories, we have
the following modified Friedmann equation
\begin{eqnarray}\label{43}
\rho=-2 F'(T)T+F(T)\,,
\end{eqnarray}
which is a curve in the plane $(H,\rho)$. Here, the prime denotes
the derivative with respect to $T$.
Moreover, the dynamics is given by
\begin{eqnarray}\label{44}
\left\{\begin{array}{ccc}
\dot{H} \Eqn{=} \frac{f(\rho)}{4(2F''(T)T+F'(T))}\\
  \\
\dot{\rho} \Eqn{=} 3Hf(\rho)\,.
\end{array}
\right. \end{eqnarray}
This implies that the dynamics of teleparallelism theories are given by
a $1$-dimensional first order equation. Note that in $F(R)$ gravity theories,
the dynamics is more complicated because it is given by a first order
differential system in $\R^3$, i.e.,
one needs three coordinates, for example, $(H,R,\rho)$.
Following this point of view, we investigate a curve of the form $\rho=G(T)$
for some function $G$. This curve could be obtained from the modified Friedmann
equation (\ref{43}) by choosing
\begin{eqnarray}\label{45}
F(T)=-\frac{\sqrt{-T}}{2}\int \frac{G(T)}{T\sqrt{-T}}dT\,.
\end{eqnarray}

What is interesting here is that for different choices of the curve
$\rho=G(T)$, we find all the kinds of singularities.
This is contrary to the case of usual LQC which was discussed in previous
sections.
For example, suppose that the function $G$ is monotonic,
the dynamical equation (\ref{44}) can be written as
\begin{eqnarray}\label{46}
\left\{\begin{array}{ccc}
\dot{H}&=& -\frac{f(\rho)}{4}(G^{-1})'(\rho)\\
\dot{\rho}&=& 3Hf(\rho).
                                \end{array}
\right.
\end{eqnarray}
By taking a dynamic $H(t)$,
from the modified Friedmann and Raycahauduri equations
\begin{eqnarray}\label{47} \rho=G(T)\quad \mbox{and} \quad
\dot{H}=-\frac{f(\rho)}{4}(G^{-1})'(\rho)\,,
\end{eqnarray}
we build the corresponding EoS (i.e., $f(\rho)$),
which yields the dynamics $H(t)$.
As a first example,
we consider the curve $\rho=\frac{H_s^4}{T+6H_s^2}$ and the
dynamics $H(t)=H_se^{-H_0(t_s-t)}$. The corresponding EoS is given by
the function
\begin{eqnarray}\label{48}
f(\rho)=\frac{4H_0\rho^2}{H_s^3}\sqrt{1-\frac{H_s^2}{6\rho}}\,.
\end{eqnarray}
In this case, when $t\rightarrow t_s$, we find $H\rightarrow H_s$,
$\rho\rightarrow \infty$ and $|P|\rightarrow\infty$, that is,
a Big Freeze occurs.

There another interesting example exists.
If we examine the curve $\rho=\rho_s\left(1- \frac{H_0^4}{H_0^4+T^2}\right)$
and the dynamics $H(t)=H_0e^{H_1t}$ with $H_0, H_1>0$,
then the corresponding EoS is given by the function
\begin{eqnarray}\label{49}
f(\rho)=\frac{4\sqrt{2}H_1}{\sqrt{3}H_0}\frac{\rho^{3/4}(\rho_s-\rho)^{5/4}}{\rho_s}\,,
\end{eqnarray}
which leads to a kind of LR, where $\rho$ does not diverge,
because it satisfies $\rho\rightarrow \rho_s$ when $t\rightarrow \infty$.
This is different to the case of third section.


\section{Conclusions}

We have explored the future singularities in LQC.
It has been shown that holonomy corrections in
LQC lead to $\rho^2$ correction term
in the Friedmann equation and,
as shown numerically in early works,
eventually the Big Rip singularity
can be removed.
In addition,
we have examined whether other kinds of future singularities
including Type II and Type IV singularities can be cured or not in LQC,
provided that the universe is filled by a barotropic fluid with the state
equation $P=-\rho-f(\rho)$.
Indeed, the Friedmann equation moves in an anti-clockwise sense along an
ellipse in the plane $(H,\rho)$. This prevents Rip singularities, but
sudden singularizes could survive provided that $f$ diverges at some energy
density smaller than the critical one.

Furthermore, by generalizing the above procedures used in LQC,
we have studied the future singularizes in $F(T)$ gravity.
It has been illustrated that from a mathematical viewpoint,
this kind of theories could be understood
as theories modeled by a one dimensional first order differential equation.
This implies that the study on this theory is easier than that
in modified gravity where the dynamics of the universe is modeled by a three
dimensional first order differential system.


We also mention that for other models
in ordinary $F(T)$ gravity, namely, not in the framework of LQC,
the Type I and Type IV singularities
can eventually appear in the finite time limit for a power-law form of $F(T)$.
Moreover, the LR and PL scenarios can be realized
for specific power-law type models of $F(T)$ gravity~\cite{bmno12}.
Accordingly, the features of future singularities occurring in $F(T)$ gravity
in the context of LQC would be different from those of other models
in  $F(T)$ gravity. In particular, it is very interesting that all
the kinds of future singularities as well as Little Rip (infinite-time)
cosmology are possible in the framework of $F(T)$ gravity.

Finally, it would be interesting to study the perturbations in LQC by
following the recent approach developed in Ref.~\cite{bct11}.
The point is to compare the time required for the disintegration of bound
objects in the theories with Rip singularity with the time for the future
decay of cosmological perturbations which may be shorter
(see~\cite{Astashenok:2012iy}).
This will be discussed elsewhere.

\section*{Acknowledgments}

J. D H. would like to thank Professor Jaume Amoros for his explanations
about the geometry behind teleparallel theories.
S.D.O. would like to appreciate very kind hospitality and support
at Eurasian National University.
K.B. sincerely acknowledges
the very kind and warm hospitality
at National Center for Theoretical Sciences and
National Tsing Hua University very much, where a part of this work has been
executed.
This work was
supported in part by
MINECO
 (Spain), project MTM2011-27739-C04-01,
FIS2010-15640, and
AGAUR (Generalitat de Ca\-ta\-lu\-nya), contract 2009SGR-345
(J.D H. and S.D.O.).


\end{document}